\documentclass[pre,preprint,showpacs]{revtex4}
\usepackage[latin1]{inputenc}
\usepackage{graphicx}
\usepackage{bm}

\begin{document}

\title{Thermal Segregation Beyond Navier-Stokes}
\author{J. Javier Brey and Nagi Khalil}
\affiliation{FÝsica Te¾rica, Universidad de Sevilla, Apartado de Correos 1065, E-41080
Sevilla, Spain}
\author{James W. Dufty}
\affiliation{Department of Physics, University of Florida, Gainesville, FL 32611, USA}

\begin{abstract}
A dilute suspension of impurities in a low density gas is described by the
Boltzmann and Boltzman-Lorentz kinetic theory. Scaling forms for the species
distribution functions allow an exact determination of the hydrodynamic
fields, without restriction to small thermal gradients or Navier-Stokes
hydrodynamics. The thermal diffusion factor characterizing sedimentation is
identified in terms of collision integrals as functions of the mechanical
properties of the particles and the temperature gradient. An evaluation of
the collision integrals using Sonine polynomial approximations is discussed.
Conditions for segregation both along and opposite the temperature gradient
are found, in contrast to the Navier-Stokes description for which no
segregation occurs.
\end{abstract}

\date{\today }
\maketitle

\section{Introduction}

\label{sec1}Consider a granular mixture of two mechanically different
species in a steady state with number densities $n_{0}({\bm r)}$ and $n({\bm %
r)}$, respectively. One component is dilute with respect to the other, $%
n_{0}({\bm r)/}n(\mathbf{r)}<<1$, such that this component has
negligible effect on the host gas. Moreover, the latter is at sufficiently
low density that the granular Boltzmann kinetic theory applies for its
intra-species collisions. The dilute component has negligible intra-species
collisions and its collisions with the host gas are described by the
granular Boltzmann-Lorentz kinetic theory \cite{RydL77}. The objective here
is to provide an exact description of segregation induced by a temperature
gradient in this context. The motivation is the description some years ago
of an exact solution to the Boltzmann equation for a steady state with
constant temperature gradient \cite{BreyFourier,BKyR09}. That analysis is
extended here to include the presence of the dilute component with a
complementary description of the exact solution to the Boltzmann-Lorentz
equation. Since there is no limitation on the size of the temperature
gradient, the results given here extend previous results on thermal
segregation obtained from the Navier-Stokes equation restricted to small
gradients \cite{Garzo06}. For the dilute conditions considered here, and the
absence of gravity, no segregation occurs at Navier-Stokes order in contrast
to the results obtained here.

The particles of the dilute component will be referred to as the
\textquotedblleft impurities\textquotedblright. The hydrodynamic fields
obtained for the host gas are zero flow velocity, constant temperature
gradient in the $x$ direction, $dT(x)/dx=\theta $ , and a constant uniform
pressure $p=n(x)T(x)$. The impurities have a temperature profile $T_{0}(x)$
proportional to the host temperature $T_{0}(x)=\gamma T(x)$, and a
non-trivial density $n_{0}(x)$ expressed in terms of the host temperature
field. In the dilute limit, the concentrations are $\rho _{0}(x)\simeq
n_{0}(x)/n(x)$ and $\rho (x)=1-\rho _{0}(x)$. They have the relationship $%
d\rho _{0}/dx=-d\rho /dx$ so any spatial variation of $\rho _{0}(x)$ implies
the opposite variation of $\rho (x)$ and segregation occurs. Here the
segregation is induced by the temperature gradient, and it is common to
introduce a thermal diffusion factor $\Lambda $ defined by%
\begin{equation}
\Lambda \frac{d\ln T(x)}{dx}=-\frac{d\ln \rho _{0}(x)}{dx}\,.  \label{1.1}
\end{equation}%
This dimensionless factor depends on the properties of the two components, $%
\Lambda =\Lambda (\alpha ,\alpha _{0},\sigma /\sigma _{0},m/m_{0},\theta
^{\ast })$, where $\alpha ,\alpha _{0}$ are the restitution coefficients for
the host-host and impurity-host collisions, $\sigma ,\sigma _{0}$ and $%
m,m_{0}$ are the species diameters and masses, and $\theta ^{\ast }=\theta
/p\sigma ^{d-1}$ is the dimensionless temperature gradient, $d$ being the geometrical dimension of the
system. In principle, $%
\Lambda $ can be positive or negative within this parameter space. The case $%
\Lambda =0$ implies no segregation, while $\Lambda $ positive (negative)
implies the impurities increase concentration against (along) the
temperature gradient. This is the thermal analogue of the Brazil nut and
reverse Brazil nut effects for gravitational segregation \cite%
{DRyC93,HKyL01,JyY02,BRyM05}.

The distribution functions for the two species are of a
``normal'' form, meaning that their dependence on space and
time occurs entirely through the hydrodynamic fields, $n(x),T(x),$ and $%
n_{0}(x)$ \cite{McL89,ByR09}. Thus, boundary conditions do not occur explicitly
but only through the determination of these fields. For example, no external
driving source is required in the kinetic equation for a stationary state,
since this is implicit in the time independence of the fields. Instead, the
stationary form of the fields is determined self-consistently from moments
of the kinetic equations. This self-consistency also determines the
temperature of the impurities as being proportioal to the host temperature, $%
T_{0}(x)=\gamma T(x)$, with $\gamma \neq 1$ in general. No reference to
hydrodynamics is made, although these moment equations are equivalent to the
balance equations forming the basis for a hydrodynamical description.

The steady state obtained occurs by establishing a gradient of the heat flux
to compensate for local energy loss due to collisional cooling. Thus it is
special to granular fluids and links the temperature gradient to the degree
of inelasticity rather than to boundary conditions. This is similar to
steady uniform shear flow where the steady state is possible due to a
balance of viscous heating and collisional cooling, such that the velocity
gradient (shear rate) is linked to the degree of inelasticity. In both
cases, the control needed to assure Navier-Stokes hydrodynamics is lost. In
the present case, smaller gradients entails smaller pressure at constant
restitution coefficient, or smaller inelasticity at constant pressure. Such
non-Newtonian steady states are a characteristic of granular flows and
segregation for such states can be qualitatively different from that from
Navier-Stokes hydrodynamics. This \ has been illustrated recently for
thermal segregation under uniform shear flow \cite{Garzo10}.

The next section defines the system and its kinetic theory description. In
section \ref{sec3} scaling forms for the distribution functions are
introduced and the implications for the hydrodynamic fields are obtained.
Three constants must be determined self-consistently. One of these, the
temperature gradient $\theta $ has been obtained in \cite{BreyFourier,BKyR09}.
Collision integrals for the other two are obtained here. The form of the
thermal diffusion factor $\Lambda $ is given in terms of these constants,
and the sign of $\Lambda $ is discussed based on approximate evaluations of
the collision integrals given in the Appendices.

\section{Kinetic theory}

\label{sec2}Consider a one component gas of $N$ smooth, inelastic hard
spheres ($d=3$) or disks ($d=2$) with diameter $\sigma $ and mass $m$ at low
density. Their distribution for position ${\bm r}$ and velocity ${\bm v}$ at
time $t$, $f({\bm
r},{\bm v},t)$, is determined from the Boltzmann equation (without external
forces) \cite{GyS95}
\begin{equation}
\left( \partial _{t}+{\bm v}\cdot \frac{\partial}{\partial {\bm r}}\right)
f=J\left[ {\bm v}|f,f\right] ,  \label{2.1}
\end{equation}%
where the collision operator $J\left[ \mathbf{v}|f,f\right] $ is%
\begin{equation}
J\left[ \mathbf{v}|f,f\right] \equiv\sigma ^{d-1}\int d{\bm v}_{1}\int d%
\widehat{\bm{\sigma }}\, \Theta (\widehat{{\bm{\sigma }}}\cdot {\bm g})(%
\widehat{\bm{\sigma }}\cdot {\bm g})\left[ \alpha ^{-2}f({\bm r},{\bm v}%
^{\prime },t)f({\bm r},{\bm v}_{1}^{\prime },t)-f({\bm r},{\bm v},t)f({\bm r}%
,{\bm v}_{1},t)\right] .  \label{2.2}
\end{equation}%
Here ${\bm g} \equiv {\bm v}-{\bm v}_{1}$ is the relative velocity of the
colliding pair, $\Theta$ is the Heaviside step function, $d \widehat{\bm %
\sigma}$ is the solid angle element about the direction of the unit vector $%
\widehat{\bm \sigma}$, and $\alpha $ is the restitution coefficient
characterizing the degree of inelasticity ($0<\alpha \leq 1$). The
velocities ${\bm v}^{\prime },{\bm v}_{1}^{\prime }$ denote the restituting
velocities for the pair ${\bm v}, {\bm v}_{1}$,

\begin{equation}
{\bm v}^{\prime }={\bm v}-\frac{1+\alpha ^{-1}}{2} (\widehat{{\bm{\sigma }}}%
\cdot \mathbf{g})\widehat{{\bm{\sigma }}},\quad {\bm v}_{1}^{\prime }={\bm v}%
_{1}+\frac{ 1+\alpha ^{-1}}{2} (\widehat{{\bm{\sigma }}}\cdot {\bm g})%
\widehat{\bm{\sigma }}.  \label{2.3}
\end{equation}

Now consider $M$ additional impurity particles in this gas, all the same but
mechanically different from the fluid particles. For $M \ll N$, the primary
collisions for the impurity particles are with the host gas particles, and
impurity-impurity collisions and effects of the impurities on the gas
distribution function $f$ can be neglected. The distribution function for
the impurities, $F({\bm r},{\bm v}_{0},t)$, is governed by the corresponding
Boltzmann-Lorentz equation,
\begin{equation}
\left( \partial _{t}+{\bm v}_{0} \cdot \frac{\partial}{\partial {\bm r}}%
\right) F=I\left[ {\bm v}_{0}|F,f\right] ,  \label{2.4}
\end{equation}%
where the operator $I\left[ {\bm v}_{0}|F,f\right] $ describes changes in $F$
due to binary collisions between the impurity and gas particles,
\begin{equation}
I\left[ {\bm v}_{0}|F,f\right] \equiv \overline{\sigma}^{d-1}\int d{\bm v}%
_{1}\int d\widehat{\bm{\sigma }}\, \Theta (\widehat{{\bm{\sigma }}}\cdot {%
\bm g}_{01})(\widehat{\bm{\sigma }}\cdot {\bm g}_{01})\left[ \alpha
_{0}^{-2}F({\bm r},{\bm v}_{0}^{\prime },t)f({\bm r},{\bm v}_{1}^{\prime
},t)-F({\bm r},{\bm v}_{0},t)f({\bm r},{\bm v}_{1},t)\right] ,  \label{2.5}
\end{equation}%
${\bm g}_{01} \equiv {\bm v}_{0}-{\bm v}_{1}$. The restituting velocities ${%
\bm v}_{0}^{\prime },{\bm v}_{1}^{\prime }$ in this case are%
\begin{equation}
{\bm v}_{0}^{\prime }={\bm v}_{0}-\frac{m \left( 1+\alpha _{0}^{-1}\right)}{%
m+m_{0}} (\widehat{{\bm{\sigma }}}\cdot \mathbf{g}_{01})\widehat{{\bm{\sigma
}}},\quad {\bm v}_{1}^{\prime }={\bm v}_{1}+\frac{m_{0}\left( 1+\alpha
_{0}^{-1}\right)}{m+m_{0}} (\widehat{{\bm{\sigma }}}\cdot {\bm g}_{01})%
\widehat{\bm{\sigma }}.  \label{2.6}
\end{equation}%
In the above expressions, $\overline{\sigma} \equiv (\sigma +\sigma_{0})/2$,
and $\sigma_{0}$, $m_{0}$, and $\alpha _{0}$ are the hard sphere diameter,
mass, and restitution coefficient for the impurity particles, respectively.

The macroscopic state of this system is described by the fluid number
density $n({\bm r},t)$, temperature $T({\bm r},t)$, and flow velocity ${\bm u%
}({\bm r},t)$, defined in terms of the distribution function by%
\begin{equation}
\left(
\begin{array}{c}
n(\mathbf{r},t) \\
\frac{d}{2}n({\bm r},t)T({\bm r},t) \\
n({\bm r},t){\bm u}({\bm r},t)%
\end{array}%
\right) \equiv \int d{\bm v}\left(
\begin{array}{c}
1 \\
\frac{1}{2}mV^{2} \\
\mathbf{v}%
\end{array}%
\right) f({\bm r},{\bm v},t),  \label{2.7}
\end{equation}%
with ${\bm V}({\bm r},t) \equiv {\bm v}-{\bm u}({\bm r},t)$. It is
convenient to introduce corresponding fields for a macroscopic description
of the impurity particles,%
\begin{equation}
\left(
\begin{array}{c}
n_{0}({\bm r},t) \\
\frac{d}{2}n_{0}({\bm r},t)T_{0}({\bm r},t) \\
\mathbf{j}_{0}({\bm r},t)%
\end{array}%
\right) \equiv \int d\mathbf{v}_{0}\left(
\begin{array}{c}
1 \\
\frac{1}{2}mV_{0}^{2} \\
{\bm v}_{0}%
\end{array}%
\right) F({\bm r},{\bm v}_{0}),  \label{2.8}
\end{equation}%
with ${\bm V}_{0}({\bm r},t)={\bm v}_{0}-\mathbf{u}_{0}({\bm r}_{0},t).$
Instead of an impurity velocity, the more usual number flux notation ${\bm j}%
_{0}=n_{0} {\bm u}_{0}$ has been used.

\section{Scaling solutions}

\label{sec3}In reference \cite{BreyFourier}, a solution to the Boltzmann
equation was described for the special case of a scaling form in terms of
the hydrodynamic variables,
\begin{equation}
f(x,{\bm v})=n(x)\left[ \frac{m}{2T(x)}\right] ^{d/2}\phi \left( {\bm c}%
\right) ,\quad {\bm c}\equiv \left[ \frac{m}{2T(x)}\right] ^{1/2}{\bm v.}
\label{3.1}
\end{equation}%
Such a solution, where the space and time dependence of the distribution
function occurs only through the hydrodynamic fields, is called
\textquotedblleft normal\textquotedblright . The definitions of the fields
in (\ref{2.7}), and the choice of ${\bm u}=0$ give the self-consistency
conditions on $\phi \left( {\bm c},\right) $
\begin{equation}
\left(
\begin{array}{c}
1 \\
\frac{d}{2} \\
0%
\end{array}%
\right) =\int d\mathbf{c}\left(
\begin{array}{c}
1 \\
c^{2} \\
\mathbf{c}%
\end{array}%
\right) \phi \left( \mathbf{c}\right) .  \label{3.2}
\end{equation}%
Here, a similar scaling solution for the impurities is sought,
\begin{equation}
F(x,{\bm v}_{0})=n_{0}(x)\left[ \frac{m_{0}}{2T_{0}(x)}\right] ^{d/2}\Phi
\left( {\bm c}_{0}\right) ,\quad {\bm c}_{0}=\left[ \frac{m_{0}}{2T_{0}(x)}%
\right] ^{1/2}{\bm v}_{0}.  \label{3.3}
\end{equation}%
The definitions (\ref{2.8}) then give the conditions on $\Phi $,
\begin{equation}
\left(
\begin{array}{c}
1 \\
\frac{d}{2}+\frac{m_{0}j_{0}^{2}}{2T_{0}n_{0}^{2}} \\
\frac{1}{n_{0}}(\frac{m_{0}}{2T_{0}})^{1/2}{\bm j}_{0}%
\end{array}%
\right) \equiv \int d{\bm c}_{0}\left(
\begin{array}{c}
1 \\
c_{0}^{2} \\
{\bm c}_{0}%
\end{array}%
\right) \Phi \left( {\bm c}_{0}\right) .  \label{3.4}
\end{equation}%
In order for (\ref{3.3}) to be \textquotedblleft normal\textquotedblright ,
it should depend only on the hydrodynamic fields for the gas and impurities,
i.e., on $n_{0}(x)$, $n(x)$, and $T(x)$. Dimensional analysis then requires
that $T_{0}(x)$ must be proportional to $T(x)$,
\begin{equation}
T_{0}(x)=\gamma T(x).  \label{3.5}
\end{equation}%
The constant $\gamma $ must be determined in course of solving the kinetic
equation (as discussed below). Further comments on the implications of
normal solutions is provided in the last section.

In terms of these scaling solutions and dimensionless velocity variables,
the Boltzmann and Boltzmann-Lorentz equations become%
\begin{equation}
c_{x}\left\{ \frac{1}{n\sigma ^{d-1}}\frac{d\ln n}{dx}\phi \left( {\bm c}%
\right) -\frac{1}{2n\sigma ^{d-1}}\frac{d\ln T}{dx}\frac{\partial }{\partial
{\bm c}}\cdot \left[ {\bm c}\phi \left( {\bm c}\right) \right] \right\} =%
\mathcal{J}\left[ {\bm c}|\phi ,\phi \right] ,  \label{3.6}
\end{equation}%
\begin{equation}
c_{0x}\left\{ \frac{1}{n\overline{\sigma }^{d-1}}\frac{d\ln n_{0}}{dx}\Phi
\left( {\bm c}_{0}\right) -\frac{1}{2n\overline{\sigma }^{d-1}}\frac{d\ln T}{%
dx}\frac{\partial }{\partial {\bm c}_{0}}\cdot \left[ {\bm c}_{0}\Phi \left(
{\bm c}_{0}\right) \right] \right\} =\mathcal{I}\left[ {\bm c}_{0}|\Phi
,\phi \right] ,  \label{3.7}
\end{equation}%
with the dimensionless collision operators%
\begin{equation}
\mathcal{J}\left[ {\bm c}|\phi ,\phi \right] \equiv \int d{\bm c}_{1}\int d%
\widehat{\bm{\sigma }}\,\Theta (\widehat{{\bm{\sigma }}}\cdot {\bm w})(%
\widehat{\bm{\sigma }}\cdot {\bm w})\left[ \alpha ^{-2}\phi \left( {\bm c}%
^{\prime }\right) \phi \left( {\bm c}_{1}^{\prime }\right) -\phi \left( {\bm %
c}\right) \phi \left( {\bm c}_{1}\right) \right] ,  \label{3.8}
\end{equation}%
\begin{equation}
\mathcal{I}\left[ {\bm c}_{0}|\Phi ,\phi \right] =\int d{\bm c}_{1}\int d%
\widehat{\bm{\sigma }}\,\Theta (\widehat{{\bm{\sigma }}}\cdot {\bm w}_{0})(%
\widehat{\bm{\sigma }}\cdot {\bm w}_{0})\times \left[ \alpha _{0}^{-2}\Phi
\left( {\bm c}_{0}^{\prime }\right) \phi \left( {\bm c}_{1}^{\prime }\right)
-\Phi \left( {\bm c}_{0}\right) \phi \left( {\bm c}_{1}\right) \right] .
\label{3.9}
\end{equation}%
The relative velocities ${\bm w}$ and ${\bm w}_{0}$ are now
\begin{equation}
{\bm w}\equiv {\bm c}-{\bm c}_{1},\quad {\bm w}_{0}\equiv {\bm c}_{0}-\left(
\frac{m_{0}}{m\gamma }\right) ^{1/2}{\bm c}.  \label{3.10}
\end{equation}%
The expressions of the dimensionless restituting velocities in Eq.\ (\ref%
{3.9}) are given in Eq.\ (\ref{a.3}). Since the right sides of Eqs.\ (\ref%
{3.6}) and (\ref{3.7}) are independent of $x$, the left sides must be as
well. This will be true if the hydrodynamic fields $n(x)$, $n_{0}(x)$, and $%
T(x)$ satisfy the equations%
\begin{equation}
\frac{1}{n\sigma ^{d-1}}\frac{d\ln n}{dx}=A,\quad \frac{1}{n\sigma ^{d-1}}%
\frac{d\ln T}{dx}=B,\quad \frac{1}{n\overline{\sigma }^{d-1}}\frac{d\ln
n_{0}(x)}{dx}=C,  \label{3.11}
\end{equation}%
where $A$, $B$, and $C$ are constants. The constants $A$ and $B$ are
determined by taking moments of the Boltzmann equation (\ref{3.6}). Namely,
multiplication of the equation by $1$, $c_{x}$, and $c^{2}$, and integration
over ${\bm c}$ yields
\begin{equation}
\left( A+\frac{B}{2}\right) \int d{\bm c}\,c_{x}\phi \left( {\bm c}\right)
=0,\quad \left( A+B\right) \int d{\bm c}\,c_{x}^{2}\phi \left( {\bm c}%
\right) =0,  \label{3.12}
\end{equation}%
\begin{equation}
\left( A+\frac{3B}{2}\right) \int d{\bm c}\,c^{2}c_{x}\phi \left( {\bm c}%
\right) =\int d{\bm c}\,c^{2}\mathcal{J}\left[ {\bm c}|\phi ,\phi \right] .
\label{3.13}
\end{equation}%
The zeroes on the right sides of (\ref{3.12}) result from conservation of
particle number and momentum by the collision operator. The first equation
of (\ref{3.12}) is satisfied because of conditions (\ref{3.2}) required on $%
\phi ({\bm c})$, while the second equation gives $A=-B$. Finally,
Eq.\thinspace\ (\ref{3.13}) determines $B$,
\begin{equation}
B=\frac{2\int d{\bm c}\,c^{2}\mathcal{J}\left[ {\bm c}|\phi ,\phi \right] }{%
\int d{\bm c}\,c^{2}c_{x}\phi \left( {\bm c}\right) }.  \label{3.15}
\end{equation}%
The exact hydrodynamic fields for the gas are now given exactly by ${\bm u}%
=0 $ and%
\begin{equation}
\frac{dp}{dx}=0,\quad \frac{dT}{dx}=\theta ,  \label{3.15a}
\end{equation}%
where $p=n(x)T(x)$ is the uniform pressure, and $\theta \equiv Bp\sigma
^{d-1}$ is the constant temperature gradient.

A similar analysis applies for the impurity constants $C$ and $\gamma $.
Taking moments of the Boltzmann-Lorentz equation (\ref{3.7}) with respect to
$1,c_{0x}\mathbf{,}$ and $c_{0}^{2}$ gives%
\begin{equation}
\left( C+\frac{\sigma ^{d-1}}{2\overline{\sigma}^{d-1}}B\right) j_{0x}=0,
\label{3.16}
\end{equation}%
\begin{equation}
C+\frac{\sigma ^{d-1}}{\overline{\sigma}^{d-1}}B=\frac{\int d{\bm c}_{0}\,
c_{0x}\mathcal{I}\left[ {\bm c}_{0}|\Phi ,\phi \right] }{\int d{\bm c}%
\,_{0}c_{0x}^{2}\Phi \left( {\bm c}_{0}\right) }\, ,  \label{3.17}
\end{equation}%
\begin{equation}
C+\frac{3\sigma ^{d-1}}{2\overline{\sigma}^{d-1}}B=\frac{\int d{\bm c}_{0}\,
c_{0}^{2}\mathcal{I}\left[ {\bm c}_{0}|\Phi ,\phi \right] }{\int d{\bm c}%
_{0} c_{0}^{2}c_{0x}\Phi ({\bm c}_{0})}\, .  \label{3.18}
\end{equation}%
The right hand sides of Eqs.\ (\ref{3.17}) and (\ref{3.18}) depend on $%
\gamma $ explicitly through ${\bm w}_{0}$ (see (\ref{3.10})) and implicitly
on both $\gamma$ and $C$ through $\Phi $. Since $B$ is known independently
from Eq.\ (\ref{3.15}), the two unknowns $\gamma$ and $C$ are determined by
Eqs.\ (\ref{3.17}) and (\ref{3.18}). Equation (\ref{3.16}) has two
solutions, $j_{0x}=0$ and $C=-\sigma^{d-1}B/2 \overline{\sigma}^{d-1} $. The
latter gives an additional equation for $\gamma$ and $C$ and the problem is
overdetermined. Probably, this choice is not consistent with the assumption (%
\ref{3.3}). Here, it is assumed that the boundary conditions enforce the
choice $j_{0x}=0$.

In summary, the description of the gas and impurities is completely
specified by the kinetic equations for $\phi \left( {\bm c}\right) $ and $%
\Phi \left( {\bm c}_{0}\right) $,
\begin{equation}
-Bc_{x}\left\{ \phi \left( {\bm c}\right) +\frac{1}{2}\frac{\partial }{%
\partial {\bm c}}\cdot \left[ {\bm c}\phi \left( {\bm c}\right) \right]
\right\} =\mathcal{J}\left[ {\bm c}|\phi ,\phi \right] ,  \label{3.19}
\end{equation}%
\begin{equation}
c_{0x}\left\{ C\Phi \left( {\bm c}_{0}\right) -\frac{\sigma ^{d-1}B}{2%
\overline{\sigma }^{d-1}}\frac{\partial }{\partial {\bm c}_{0}}\cdot \left[ {%
\bm c}_{0}\Phi \left( {\bm c}_{0}\right) \right] \right\} =\mathcal{I}\left[
{\bm c}_{0}|\Phi ,\phi \right] ,  \label{3.20}
\end{equation}%
and the constants $B$, $C$, and $\gamma $ are determined self-consistently
from Eqs. (\ref{3.15}), (\ref{3.17}), and (\ref{3.18}). The corresponding
collision integrals are further simplified in Appendix \ref{App1}. The
hydrodynamic fields have the simple spatial forms%
\begin{equation}
T(x)=T(0)+\theta x,\quad n(x)=\frac{p}{T(0)+\theta x}\,,  \label{3.21}
\end{equation}%
\begin{equation}
T_{0}(x)=T_{0}(0)+\gamma \theta x,\quad n_{0}(x)=n_{0}(0)\left[ 1+\frac{%
\theta x}{T(0)}\right] ^{\overline{\sigma }^{d-1}C/\sigma ^{d-1}B}\,.
\label{3.22}
\end{equation}

\section{Segregation}

\label{sec4}The segregation of impurity particles relative to the host gas
is described by the inhomogeneity of the composition $\rho
_{0}(x) \simeq n_{0}(x)/n(x)$, which follows from (\ref{3.21}) and (\ref{3.22})%
\begin{equation}
\rho _{0}\left( x\right) =\frac{n_{0}(0)}{n(0)}\left[ 1+\frac{\theta }{T(0)}x%
\right] ^{1+\overline{\sigma }^{d-1}C/\sigma ^{d-1}B}.  \label{4.1}
\end{equation}%
The thermal diffusion factor of (\ref{1.1}) is therefore%
\begin{equation}
\Lambda =-\left( 1+\frac{\overline{\sigma }^{d-1}C}{\sigma ^{d-1}B}\right) =-%
\frac{\overline{\sigma }^{d-1}}{2\sigma ^{d-1}}\frac{\int d{\bm c}%
_{0}\,c_{0x}\mathcal{I}\left[ {\bm c}_{0}|\Phi ,\phi \right] }{\int d{\bm c}%
_{0}\,c_{0x}^{2}\Phi \left( {\bm c}_{0}\right) }\frac{\int d{\bm c}%
\,c^{2}c_{x}\phi \left( {\bm c}\right) }{\int d{\bm c}\,c^{2}\mathcal{J}%
\left[ {\bm c}|\phi ,\phi \right] }\, .  \label{4.2}
\end{equation}%
If the impurities are mechanically equivalent to the host particles, then $%
\mathcal{I}\left[ {\bm c}_{0}|\Phi ,\phi \right] =\mathcal{J}\left[ {\bm c}%
|\phi ,\phi \right] ,$ $\Phi =\phi ,$ and $\Lambda =0$, since the first
integral in the numerator of (\ref{4.2}) vanishes by conservation of
momentum.

The corresponding result for the thermal diffusion factor obtained from the
Navier-Stokes order Chapman-Enskog solutions to the Boltzmann and
Boltzmann-Lorentz equations gives $\Lambda =0$ for all values of the
parameters $\alpha ,\alpha _{0},\sigma /\sigma _{0},m/m_{0},\theta /p\sigma
^{d-1}$. If the Navier-Stokes calculation is extended to include effects of
gravity the condition becomes \cite{Garzo06}%
\begin{equation}
\left( \Lambda \frac{\partial T}{\partial x} \right) _{NS}=mg\left( \frac{T_{0}}{T}-\frac{%
m_{0}}{m}\right) .  \label{4.3}
\end{equation}%
Thus thermal segregation can occur, facilitated by gravity, and depends on
the sign of $\left( T_{0}/T-m_{0}/m\right) $ and the direction of $\partial
T / \partial x$ relative to the gravitational force. This is in sharp contrast to the
results obtained in the next section.

\section{Approximate determination of $T_{0}/T$ and $\Lambda $}

\label{sec5}To determine the coefficients $B,C,$ and $\gamma =T_{0}/T$, the
distribution functions $\phi $ and $\Phi $ are represented as truncated
Sonine polynomial expansions
\begin{eqnarray}
\phi \left( {\bm c}\right) &\simeq &\pi ^{-d/2}e^{-c^{2}}\left[
1-a_{01}\left( c^{2}-dc_{x}^{2}\right) +\left( \frac{d-1}{2}\,b_{01}+\frac{3%
}{2}b_{10}\right) c_{x}\right.  \nonumber \\
&&\left. -b_{01}c^{2}c_{x}-\left( b_{10}-b_{01}\right) c_{x}^{3}\right] ,
\label{5.1}
\end{eqnarray}%
\begin{eqnarray}
\Phi \left( {\bm c}_{0}\right) &\simeq &\pi ^{-d/2}e^{-c_{0}^{2}}\left[
1-A_{01}\left( c_{0}^{2}-dc_{0x}^{2}\right) +\left( \frac{d-1}{d}B_{01}+%
\frac{3}{2}B_{10}\right) c_{0x}\right.  \nonumber \\
&&\left. -B_{01}c_{0}^{2}c_{0x}-\left( B_{10}-B_{01}\right) c_{0x}^{3}\right]
.  \label{5.2}
\end{eqnarray}%
The method for determining the coefficients in these expansions is described
in \cite{BreyFourier} and summarized for the case here in Appendix \ref{App2}%
. The numerical solutions for the case of a two-dimensional system ($d=2$)
with $m=m_{0}$ and $\sigma =\sigma _{0}=\overline{\sigma }$ are shown in
Figs. \ref{fig7}-\ref{fig9} as a function of $\alpha $ for
several values of $\alpha _{0}$. An important general feature is that all
coefficients in (\ref{5.1}) and (\ref{5.2}) vanish as $\alpha \rightarrow 1$%
. Thus the non-uniform steady state described here exists only as a
consequence of the inelasticity of the host gas. Further comment on this is
given in the last section below.

\begin{figure}
\includegraphics[scale=0.8,angle=0]{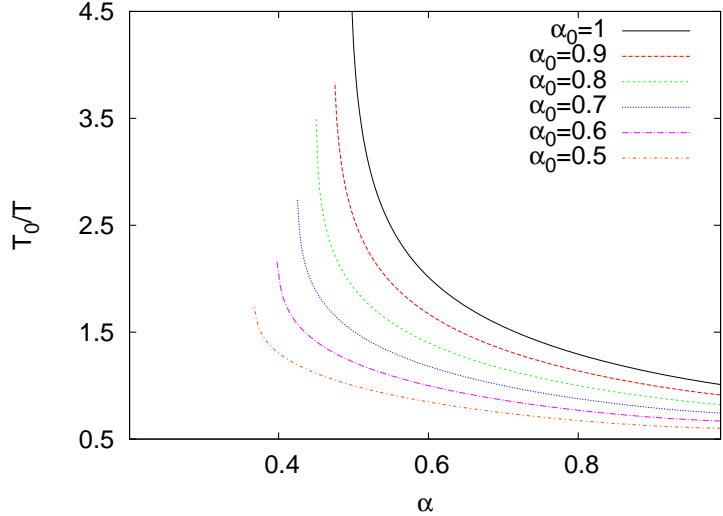}
\caption{Temperature of the impurity $T_{0}$ divided by the temperature of the hot gas $T$ as
a function of the coefficient of normal restitution of the gas particles $\alpha$, for several values
of the restitution coefficient for collisions between the gas particles and the impurities, $\alpha_{0}$, as
indicated in the insert. In all cases, $d=2$,  $m_{0}=m$, and   $\sigma =\sigma _{0}=\overline{%
\sigma }$.} \label{fig1}
\end{figure}

In the following, attention is restricted to $\sigma =\sigma _{0}=\overline{%
\sigma }$ and $d=2$ for several values of $m_{0}/m$, $\alpha $, and $\alpha
_{0}$. It is well-established that different species of granular mixtures
have different partial temperatures, even in their homogeneous cooling state
(i.e., equipartition of energy does not occur) \cite{FyM02,DRyC93,GyD99}. Figures \ref{fig1} and \ref%
{fig2} show the behavior of $T_{0}/T$ for $m_{0}/m=1$ and $2$, respectively,
as a function of $\alpha $ for several values of $\alpha _{0}$. The common
feature is increasing $T_{0}/T$ with decreasing $\alpha $, increasing $%
\alpha _{0}$, and decreasing $m_{0}/m$. Figure \ref{fig3} shows a broader
range of $m_{0}/m$. Even for the relatively weak dissipation values of this
figure, it is clear that the largest values of $T_{0}/T$ occur for small
mass ratio, maximum host dissipation, and weakest impurity dissipation.

\begin{figure}
\includegraphics[scale=0.8,angle=0]{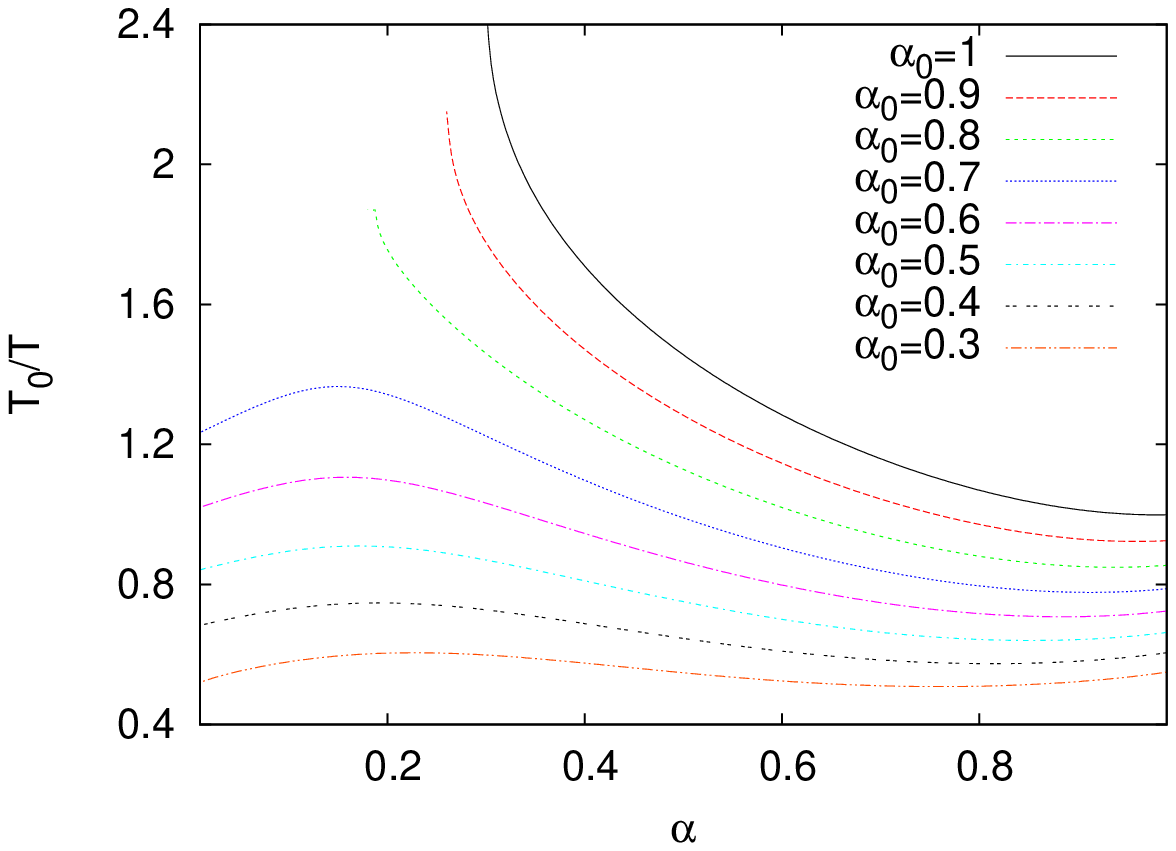}
\caption{The same as in Fig. \protect{\ref{fig1}}, but now $m_{0}/m=2$.} \label{fig2}
\end{figure}

\begin{figure}
\includegraphics[scale=0.8,angle=0]{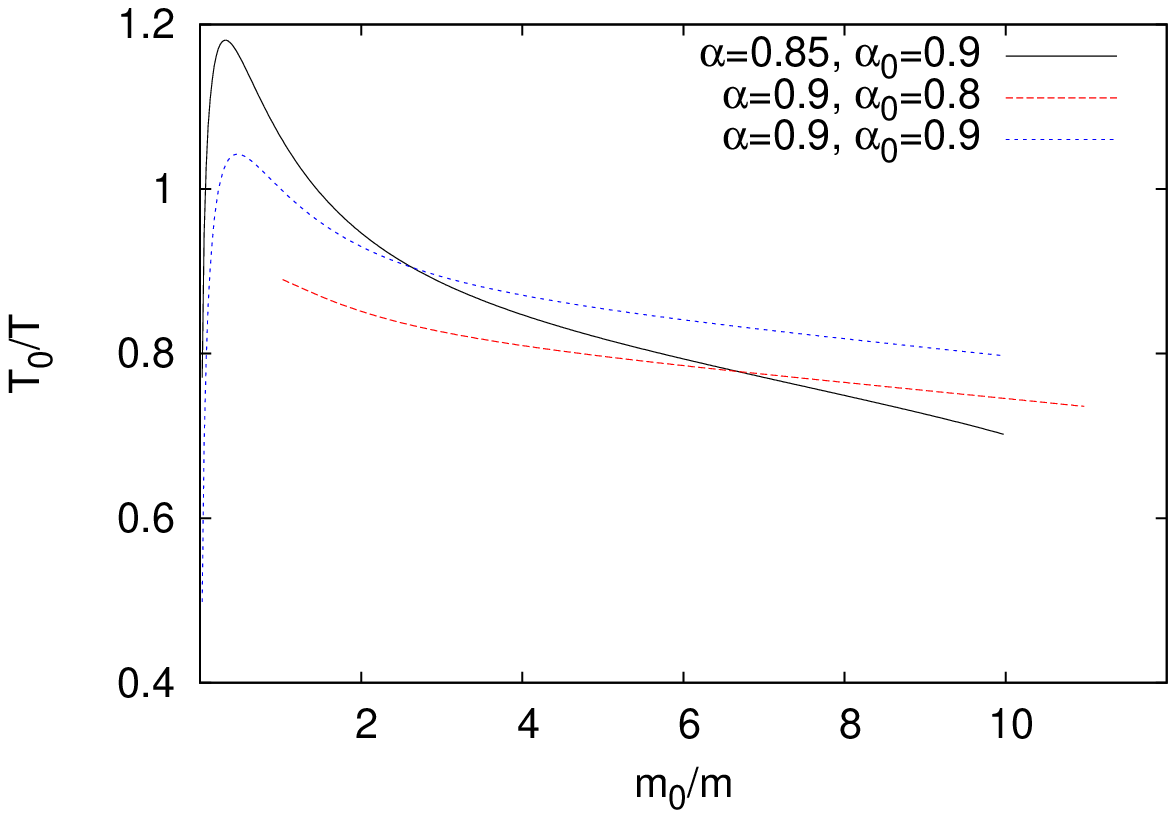}
\caption{Temperature of the impurity $T_{0}$ divided by the temperature of the hot gas $T$ as a function of
the mass ratio $m_{0}/m$ for several values of the coefficients of normal restitution $\alpha$ and $\alpha_{0}$, as
indicated in the insert. In all cases, it is $d=2$ and $\sigma = \sigma_{0}.$} \label{fig3}
\end{figure}

\begin{figure}
\includegraphics[scale=0.8,angle=0]{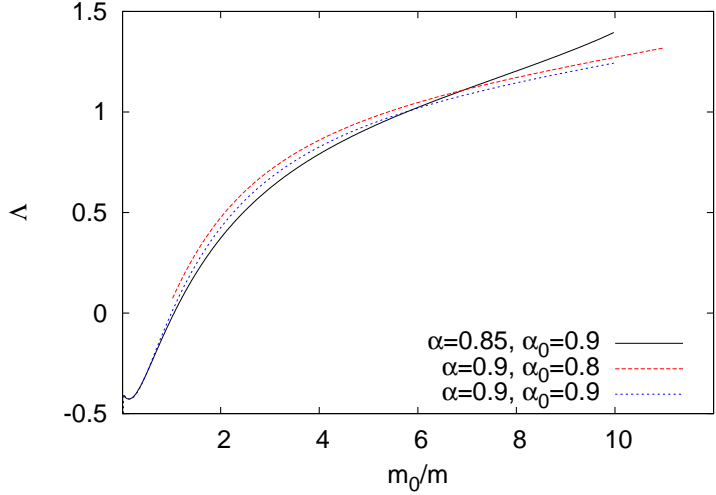}
\caption{Dimensionless thermal diffusion factor $\Lambda$ as a function of the mass ratio $m_{0}/m$ for the
same system as in Fig. \protect{\ref{fig3}}. } \label{fig4}
\end{figure}

\begin{figure}
\includegraphics[scale=0.8,angle=0]{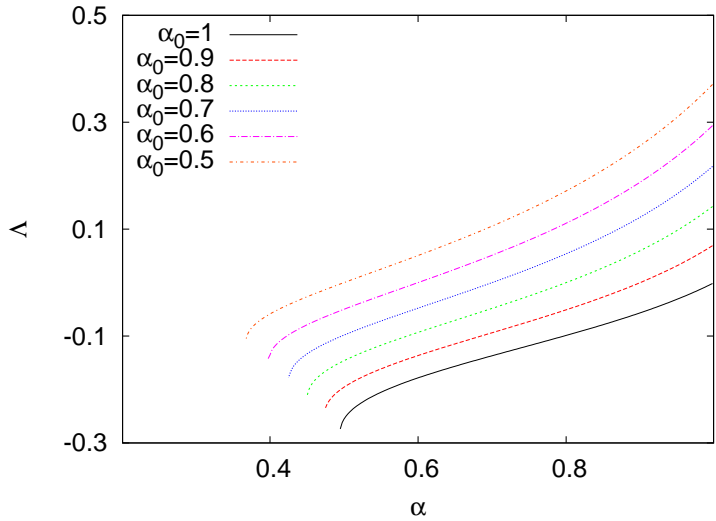}
\caption{Dimensionless thermal diffusion factor $\Lambda$ as
a function of the coefficient of normal restitution of the gas particles $\alpha$, for several values
of the restitution coefficient for collisions between the gas particles and the impurities, $\alpha_{0}$, as
indicated in the insert. In all cases, $d=2$,  $m_{0}=m$, and   $\sigma =\sigma _{0}=\overline{%
\sigma }$.} \label{fig5}
\end{figure}

\begin{figure}
\includegraphics[scale=0.7,angle=0]{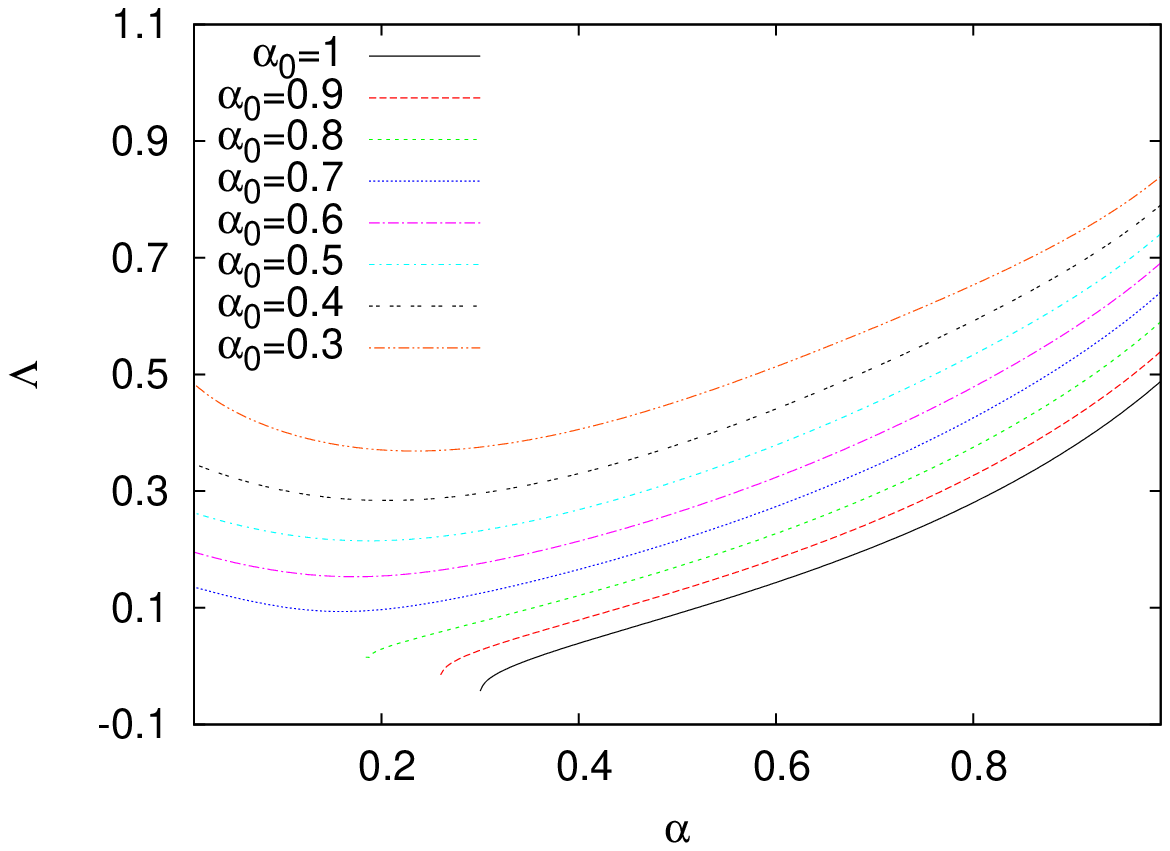}
\caption{The same as in Fig. \protect{\ref{fig5}}, but now $m_{0}/m=2$. } \label{fig6}
\end{figure}

The existence of segregation for the same weak dissipation values of Fig.\
\ref{fig3} is demonstrated in figure \ref{fig4}. The thermal diffusion factor $%
\Lambda $ is positive for $m_{0}/m>1$. This means that the impurity
concentration is higher at the colder part of the host fluid. This
is
similar to the host fluid density which behaves as $n=p/T$ with constant $p$%
. For smaller mass ratio, segregation goes in the opposite direction with
the impurity concentration highest in the hotter part of the host fluid.
This effect is enhanced at stronger host fluid dissipation and weaker
impurity dissipation, as illustrated in figures \ref{fig5} and \ref{fig6}
for $m_{0}/m=1$ and $2$, respectively. It is interesting to note that for $%
m_{0}/m=1$ the border between the two types of segregation, $\Lambda =0$,
occurs for $\alpha =\alpha _{0}$. Referring to figure \ref{fig4}, these
values also correspond to $T_{0}/T=1$. Similarly, for $m_{0}/m=2$ comparing %
Figs.\ \ref{fig2} and \ref{fig6}, it is seen that $\Lambda =0$ for $T_{0}/T=2$.
This limited data suggest the possibility that the segregation criterion $%
\Lambda =0$ occurs for $m_{0}/m=T_{0}/T$. Surprisingly, this is the same as
the Navier-Stokes criterion in the presence of gravity, (\ref{4.3}). Further
analysis of this potential relationship across a larger data set is
required. For larger $m_{0}/m$ it is found that $T_{0}/T\leq 1$, and only
the segregation for $\Lambda >0$ occurs.

\section{Discussion}

\label{sec6}The description of a low density granular gas with a
dilute concentration of impurities has been given in terms of
solutions to the coupled Boltzmann and Boltzmann-Lorentz kinetic
equations. These are normal solutions whose space and time
dependence are entirely specified in terms of the hydrodynamic
fields $n,n_{0}, $and $T$. The special case of a steady state in
which the host gas has a constant temperature gradient and constant
pressure, described earlier in refs. \cite{BreyFourier} and \cite{BKyR09}, has
been generalized to include a corresponding steady state of the
impurities. In this way the thermal segregation factor is identified
in terms of the constants of the hydrodynamic fields, without the
limiting approximations of small spatial gradients. The
self-consistent kinetic equations (\ref{3.19}) and (\ref{3.20})
determining these constants was solved using a low order Sonine
polynomial approximation for the velocity dependence of the host and
impurity distributions. The resulting thermal diffusion factor was
found to identify conditions for both segregation along and against
the temperature gradient. Such normal solutions are typically
constructed by the Chapman-Enskog method whose practical application
typically entails limitations to small spatial gradients, e.g.
Navier-Stokes order. Application of Navier-Stokes hydrodynamics
obtained in this way, and specialized to the steady state with
constant temperature gradient and constant pressure, leads to the
prediction of no segregation. The effects described here therefore
are due to contributions from the Chapman-Enskog method beyond the
small gradient approximation. In fact, there are no limitations on
the temperature gradient in the present analysis.

There are two important clarifications to note. First, the validity of a
normal solution both for granular and molecular gases is limited to domains
away from the initial preparation time and confining boundaries. For the
steady state considered here, this means that there is typically a boundary
layer across which the normal solution does not apply. Additional
information is then required to connect the physically specified values of
the fields or their gradients at the boundary with those values associated
with the normal solution. These are the familiar "slip" boundary conditions.
The existence of the normal solution described here for a system with finite
confinement and associated boundary layer has been demonstrated by molecular
dynamics simulation in refs.\cite{BreyFourier} and \cite{BKyR09}. Typically, the size of
the bulk interior relative to the boundary layer decreases as the
temperature gradient is increased. Investigation of this problem for a
molecular gas has demonstrated that the bulk normal solution domain still
exists beyond the Navier-Stokes limit \cite{Kim89}.

A second clarification is the special nature of the steady state described
here as being unique to a granular gas. The analysis of \cite{BreyFourier}
shows that it results from the balance of the heat flux gradient and the
cooling rate due to inelastic collisions. In the absence of the latter there
is no steady state solution of the type considered here. In contrast to
normal fluids, the gradients of such steady states are controlled by
internal processes rather than boundary sources. External control of the
gradients is therefore lost. In the present case the magnitude of the
dimensionless temperature gradient $\theta /p\sigma ^{d-1}=B\left( \alpha
\right) $ monotonically decreases to zero as $\alpha \rightarrow 1$,
vanishing in the elastic limit. Consequently, for example, it is not
possible for the Navier-Stokes to apply here for strong dissipation.

\section{Acknowledgments}

The research of JJB and NK has been partially supported by the Ministerio de Educaci%
¾n y Ciencia (Spain) through Grant No. FIS2008-01339 (partially financed by
FEDER funds).

\appendix

\section{Reduction of collision integrals}

\label{App1}The Boltzmann collision integral appearing on the right side of
Eq.\ (\ref{3.15}) is simplified further in \cite{BreyFourier}, with the
result%
\begin{equation}
B=- \frac{\left( 1-\alpha ^{2}\right) \pi^{(d-1)/2} }{2 \Gamma \left( \frac{%
d+3}{2} \right)} \frac{\int d{\bm c} \int d{\bm c}_{1}\left\vert {\bm c}-{%
\bm c}_{1}\right\vert^{3} \phi \left( {\bm c}\right) \phi \left( {\bm c}%
_{1}\right) }{\int d{\bm c}\, c^{2}c_{x}\phi \left( {\bm c}\right) }.
\label{a.1}
\end{equation}%
The Boltzmann-Lorentz collision integrals can be simplified in a similar
way. Consider first the collision integral appearing in Eq.\, (\ref{3.17}),
\begin{equation}
\int d{\bm c}_{0}\, c_{0x}\mathcal{I}\left[ {\bm c}_{0}|\Phi ,\phi \right]
=\int d{\bm c}_{1} \int d{\bm c}_{0}\, c_{0x}\int d\widehat{\bm{\sigma }}\,
\Theta (\widehat{{\bm{\sigma }}}\cdot {\bm w}_{0})(\widehat{\bm{\sigma }}%
\cdot {\bm w}_{0}) \left[ \alpha _{0}^{-2}\Phi \left( {\bm c}_{0}^{\prime
}\right) \phi \left( {\bm c}^{\prime }_{1}\right) -\Phi \left( {\bm c}%
_{0}\right) \phi \left( {\bm c}_{1}\right) \right] ,  \label{a.2}
\end{equation}%
where ${\bm w}_{0}$ is defined in Eq.\ (\ref{3.10}) and the dimensionless
restituting velocities following from Eq.\ (\ref{2.6}) are
\begin{equation}
{\bm c}_{0}^{\prime }={\bm c}_{0}-\frac{m}{m+m_{0}}\left( 1+\alpha
_{0}^{-1}\right) (\widehat{{\bm{\sigma }}}\cdot {\bm w}_{0})\widehat{{%
\bm{\sigma }}},\quad {\bm c}^{\prime }={\bm c}+\frac{m_{0}}{m+m_{0}}\left(
1+\alpha _{0}^{-1}\right) (\frac{m \gamma}{m_{0}})^{1/2}(\widehat{{%
\bm{\sigma }}}\cdot {\bm w}_{0})\widehat{\bm{\sigma }}.  \label{a.3}
\end{equation}
It is easily verified that
\begin{equation}
d{\bm c}^{\prime}_{1}{\bm c}_{0}^{\prime}=\alpha _{0}^{-1}d{\bm c}_{1}d{\bm c%
}_{0},\quad \widehat{{\bm{\sigma }}}\cdot \mathbf{g}_{0}=-\alpha _{0}%
\widehat{{\bm{\sigma }}}\cdot \mathbf{g}_{0}^{\prime }.  \label{a.4}
\end{equation}%
Also, Eqs.\ (\ref{a.3}) can be inverted to get the collision rule in
dimensionless units,
\begin{equation}
{\bm c}_{0}^{*}= {\bm c}_{0}-\frac{m \left( 1+\alpha _{0}\right)}{m+m_{0}} (%
\widehat{{\bm{\sigma }}}\cdot {\bm w}_{0}) \widehat{{\bm{\sigma }}},
\label{a.5}
\end{equation}%
\begin{equation}
{\bm c}^{*}_{1}={\bm c}_{1}+\frac{m_{0}\left( 1+\alpha _{0}\right)}{m+m_{0}}%
\left(\frac{m\gamma}{m_{0}}\right)^{1/2}(\widehat{{\bm{\sigma }}}\cdot {\bm w%
}_{0})\widehat{\bm{\sigma }}.  \label{a.6}
\end{equation}%
Returning to Eq.\ (\ref{a.2}), change variables in the first term of the
brackets on the right hand side to integrate over the restituting
velocities. Using the above relations, the equation becomes
\begin{eqnarray}
\int d{\bm c}_{0}\, c_{0x}\mathcal{I}\left[ {\bm c}_{0}|\Phi ,\phi \right]
&=& \int d{\bm c}_{1} \int d{\bm c}_{0}\, \Phi \left( {\bm c}_{0}\right)
\phi \left( {\bm c}_{1}\right) \int d\widehat{\bm{\sigma }}\,\Theta (%
\widehat{{\bm{\sigma }}}\cdot {\bm w}_{0})(\widehat{\bm{\sigma }}\cdot {\bm w%
}_{0})\left( c_{0x}^{* }-c_{0x}\right)  \nonumber \\
&=&- \frac{m \left( 1+\alpha _{0}\right)}{m+m_{0}} \int d{\bm c}_{1} \int d{%
\bm c}_{0}\, \Phi \left( {\bm c}_{0}\right) \phi \left( {\bm c}_{1}\right)
w_{0}w_{0x}\int d\widehat{\bm{\sigma }}\,\Theta (\widehat{{\bm{\sigma }}}%
\cdot {\bm w}_{0})(\widehat{\bm{\sigma }}\cdot \widehat{\bm w}_{0})^{3}
\nonumber \\
&=&- \frac{m\left( 1+\alpha _{0}\right)}{m+m_{0}}\frac{\pi^{(d-1)/2} }{%
\Gamma \left( \frac{d+3}{2} \right)}\int d\mathbf{c}_{1} \int d{\bm c}_{0}\,
\Phi \left( {\bm c}_{0}\right) \phi \left( {\bm c}_{1}\right) w_{0}w_{0x}.
\label{a.7}
\end{eqnarray}%
Finally (\ref{3.17}) becomes%
\begin{equation}
C+\frac{\sigma ^{d-1}}{\overline{\sigma}^{d-1}}B=- \frac{m \left( 1+\alpha
_{0}\right)}{m+m_{0}} \frac{\pi^{(d-1)/2} }{\Gamma \left( \frac{d+3}{2}
\right)}\frac{\int d{\bm c}_{1} \int d{\bm c}_{0}\, \Phi \left( {\bm c}%
_{0}\right) \phi \left( {\bm c}_{1}\right) w_{0}w_{0x}}{\int d{\bm c}_{0}\,
c_{0x}^{2}\Phi \left( {\bm c}_{0}\right) }.  \label{a.8}
\end{equation}%
The analysis of Eq.\, (\ref{3.18}) is similar with the result%
\begin{equation}
C+\frac{3\sigma ^{d-1}}{2\overline{\sigma}^{d-1}}B= \frac{m\left( 1+\alpha
_{0}\right)}{m+m_{0}} \frac{\pi^{(d-1)/2} }{\Gamma \left( \frac{d+3}{2}
\right)} \frac{\int d{\bm c}_{1} \int d{\bm c}_{0}\, \Phi \left( {\bm c}%
_{0}\right) \phi \left( {\bm c}_{1} \right) \left[ \frac{m(1+\alpha_{0})}{%
m+m_{0}}\ w_{0}^{3}- 2 w_{0} {\bm w}_{0} \cdot {\bm c}_{0} \right]}{\int d{%
\bm c}_{0}\, c_{0}^{2}c_{0x}\Phi ({\bm c}_{0})}\, .  \label{a.9}
\end{equation}

\section{Solutions to kinetic equations}

\label{App2}The solution to the kinetic equation for $\phi \left( {\bm c}%
\right) $ and the self-consistent determination of $B$ is a problem that is
independent of the impurities and can be carried out first. The method is
described in \cite{BreyFourier}. First, $B$ is given its representation as a
collision integral using Eq.\ (\ref{3.15}), so the kinetic equation (\ref%
{3.19}) becomes%
\begin{equation}
-2c_{x}\left\{ \phi \left( {\bm c}\right) +\frac{1}{2}\frac{\partial}{%
\partial {\bm c}} \cdot \left[ {\bm c}\phi \left( {\bm c}\right) \right]
\right\} \frac{\int d{\bm c}\, c^{2}\mathcal{J}\left[ {\bm c}|\phi ,\phi %
\right] }{\int d{\bm c}\, c^{2}c_{x}\phi \left( {\bm c}\right) } =\mathcal{J}%
\left[ {\bm c}|\phi ,\phi \right] .  \label{b.0}
\end{equation}%
Next $\phi \left( {\bm c}\right) $ is approximated by a truncated Sonine
polynomial expansion
\begin{eqnarray}
\phi \left( {\bm c}\right) &\simeq &\pi ^{-d/2}e^{-c^{2}}\left[
1-a_{01}\left( c^{2}-dc_{x}^{2}\right) +\left( \frac{d-1}{2}\, b_{01}+\frac{3%
}{2}b_{10}\right) c_{x}\right.  \nonumber \\
&&\left. -b_{01}c^{2}c_{x}-\left( b_{10}-b_{01}\right) c_{x}^{3}\right] .
\label{b.1}
\end{eqnarray}%
This form assures the conditions given in Eq.\ (\ref{3.2}). The coefficients
$a_{01},b_{01} $, and $b_{10}$ are then obtained from three equations
following by taking velocity moments in (\ref{b.0}). Namely, the equation is
multiplied by $c_{x}^{2},c_{x}^{3}$, and $c_{x}c^{2} $, respectively, and
afterwards integrated over ${\bm c}$. With these coefficients determined, $B$
is calculated from Eq.\ (\ref{3.15}).

To determine $\Phi \left( {\bm c}_{0}\right) $, $C$, and $\gamma $, a
similar procedure is followed. First, express $C$ as a collision integral
from Eqs.\ (\ref{3.17}) and (\ref{3.18}),
\begin{equation}
C=2\left[ \frac{3}{2}\frac{\int d{\bm c}_{0}\,c_{0x}\mathcal{I}\left[ {\bm c}%
_{0}|\Phi ,\phi \right] }{\int d{\bm c}_{0}\,c_{0x}^{2}\Phi \left( {\bm c}%
_{0}\right) }-\frac{\int d{\bm c}_{0}\,c_{0}^{2}\mathcal{I}\left[ {\bm c}%
_{0}|\Phi ,\phi \right] }{\int d{\bm c}\,_{0}c_{0}^{2}c_{0x}\Phi ({\bm c}%
_{0})}\right] ,  \label{b.2}
\end{equation}%
and use this in the kinetic equation (\ref{3.20}). Next, express $\Phi
\left( {\bm c}_{0}\right) $ as a truncated Sonine polynomial expansion
\begin{eqnarray}
\Phi \left( {\bm c}_{0}\right) &\simeq &\pi ^{-d/2}e^{-c_{0}^{2}}\left[
1-A_{01}\left( c_{0}^{2}-dc_{0x}^{2}\right) +\left( \frac{d-1}{d}B_{01}+%
\frac{3}{2}B_{10}\right) c_{0x}\right.  \nonumber \\
&&\left. -B_{01}c_{0}^{2}c_{0x}-\left( B_{10}-B_{01}\right) c_{0x}^{3}\right]
,  \label{b.3}
\end{eqnarray}%
which satisfies the conditions (\ref{3.4}) with $\mathbf{j}_{0}=0$. The
coefficients, $A_{01},B_{01}$, and $B_{10}$ are determined from three
equations obtained by taking moments of (\ref{3.20}) with respect to $%
c_{0x}^{2},c_{0x}^{3},$ and $c_{0x}c_{0}^{2}$. However, these equations also
depend on $\gamma $, so they are supplemented by an additional equation
relating the above coefficients to $\gamma $. It is obtained from a new
combination of Eqs.\ (\ref{3.17}) and (\ref{3.18})%
\begin{equation}
B=\frac{2\overline{\sigma }^{d-1}}{\sigma ^{d-1}}\left[ \frac{\int d{\bm c}%
\,_{0}c_{0}^{2}\mathcal{I}\left[ {\bm c}_{0}|\Phi ,\phi \right] }{\int d{\bm %
c}_{0}\,c_{0}^{2}c_{0x}\Phi ({\bm c}_{0})}-\frac{\int d{\bm c}\,_{0}c_{0x}%
\mathcal{I}\left[ {\bm c}_{0}|\Phi ,\phi \right] }{\int d{\bm c}%
_{0}\,c_{0x}^{2}\Phi \left( {\bm c}_{0}\right) }\right] .  \label{b.4}
\end{equation}%
Since $\phi $ and $B$ are known at this point, this gives four independent
equations for the coefficients $A_{01},B_{01},B_{10}$, and $\gamma $. With
these determined, $C$ is calculated from Eq.\ (\ref{b.2}).

\begin{figure}
\includegraphics[scale=0.8]{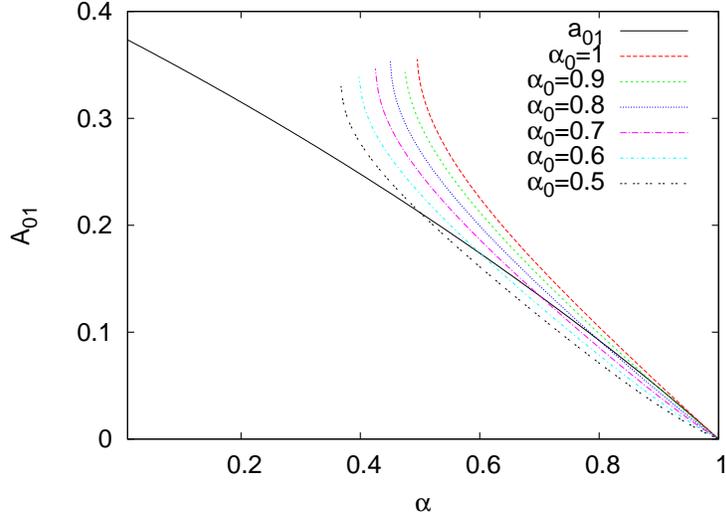}
\caption{The dimensionless parameters $a_{01}$ and $A_{01}$ as a function of
the coefficient of normal restitution of the host gas particles $\protect%
\alpha$, for several values of the coefficient of restitution for the
collisions between the gas particles and the impurities, $\protect\alpha_{0}$%
. The coefficient $a_{01}$ does not depend on the latter. The other (fixed)
parameters are $d=2$, $m_{0}=m$, and $\protect\sigma_{0}= \protect\sigma$.}
\label{fig7}
\end{figure}

\begin{figure}
\includegraphics[scale=0.8]{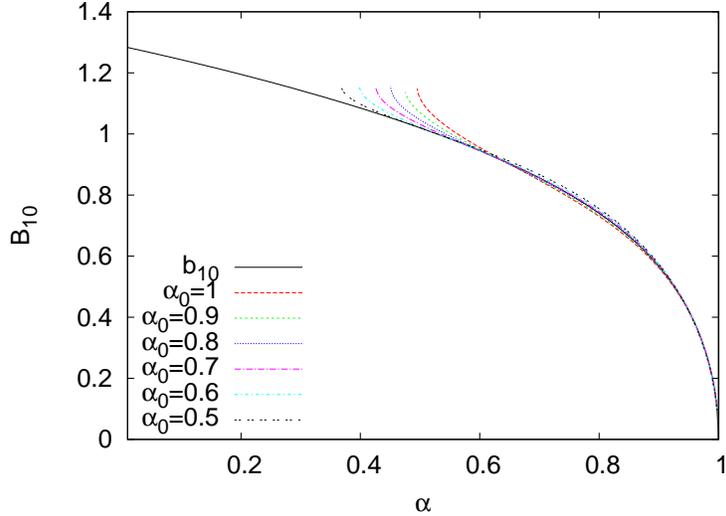}
\caption{The same as in Fig.\ {\protect\ref{fig1}} but for the coefficients $%
b_{10}$ and $B_{10}$. }
\label{fig8}
\end{figure}

\begin{figure}
\includegraphics[scale=0.8]{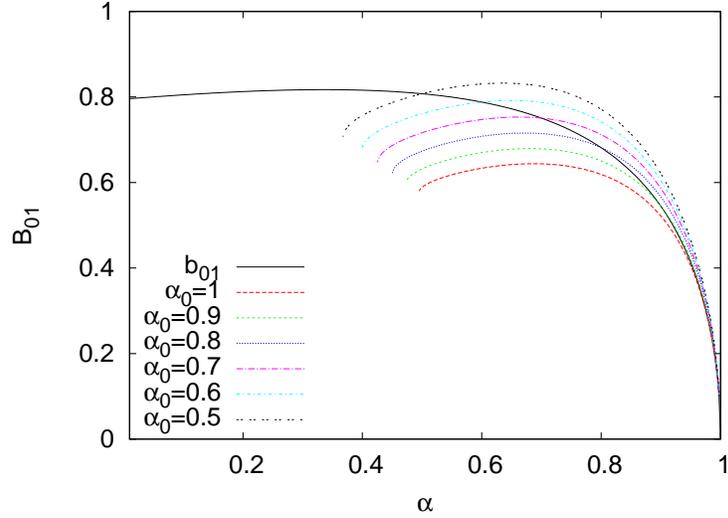}
\caption{The same as in Fig.\ {\protect\ref{fig1}}, but for $b_{01}$ and $%
B_{01}$. }
\label{fig9}
\end{figure}

In practice, the above procedure leads to highly nonlinear equations for the
coefficients. In the numerical results to be presented in the following,
only terms up to second degree in the coefficients have been kept \cite%
{BKyR09}. As an example, in Figs.\, \ref{fig7}-\ref{fig9}, the parameters
obtained for a two-dimensional system ($d=2$) with $m=m_{0}$ and $%
\sigma=\sigma_{0}= \overline{\sigma}$ are plotted as a function of $\alpha$
for several values of $\alpha_{0}$. For small values of $\alpha$, the
numerical solutions for the $B$ parameters constructed as described above
seem to disappear.

\end{document}